# Credit Cycles, Securitization, and Credit Default Swaps

Juan Ignacio Peña(*)

## Abstract

We present a limits-to-arbitrage model to study the impact of securitization, leverage and credit risk protection on the cyclicity of bank credit. In a stable bank credit situation, no cycles of credit expansion or contraction appear. Unlevered securitization together with mis-pricing of securitized assets increases lending cyclicality, favoring credit booms and busts. Leverage changes the state of affairs with respect to the simple securitization. First, the volume of real activity and banking profits increases. Second, banks sell securities when markets decline. This selling puts further pressure on falling prices. The mis-pricing of credit risk protection or securitized assets influences the real economy. Trading in these contracts reduces the amount of funding available to entrepreneurs, particularly to high-credit-risk borrowers. This trading decreases the liquidity of the securitized assets, and especially those based on investments with high credit risk.

* Corresponding author. Juan Ignacio Peña is at the Department of Business Administration, Universidad Carlos III de Madrid, c/ Madrid 126, 28903 Getafe (Madrid, Spain) ypenya@eco.uc3m.es.

# 1. Introduction

An economy with stable bank credit is in a situation without economic reasons for cycles of credit expansion or contraction. Lending cycles appear when funding by banks concentrates in specific periods of time but in other periods only scarce funding is available for real-economy projects. In good times, banks follow lax credit policies with lower credit standards, and, so negative-NPV loans increase. During recessions, when banks are burdened with many nonperforming loans, banks tighten credit standards and may refuse to give positive-NPV loans (Jimenez and Saurina, 2006). Therefore, the cyclicality of bank credit is a crucial determinant of the volatility of real activity, output and employment.

Current literature focuses on the foundations of credit cyclicality, and the crucial role of the variations in credit supply in explaining the evolution of the business cycle, see Holmström and Tirole (1997), Bernanke and Gertler (1989), Kiyotaki and Moore (1997), and Diamond and Rajan (2005), among others. But the current literature is less informative about the role played by the market for credit risk transfer as a possible reason influencing the credit cyclicality.

In this paper, we contribute to filling this gap in the literature, by studying the extent to which the trading in the most popular credit derivative instrument, the Credit Default Swap (CDS), affects the incentives of banks about lending, securitization, borrowing, and leverage. We present implications for the cyclicality of the bank credit supply and for the volatility of economic activity, by using a theoretical model based on a limits-to-arbitrage framework (Shleifer and Vishny ,1997).

The Credit Default Swap (CDS), introduced in 1997[1], has become the most popular credit derivative contract. Market size measured by gross notional amount decreases since its largest in 2007 (60,000 $ billions) and is around 9,500 $ billions in the last quarter of 2017. A CDS is a contract between two parties, the protection buyer and a protection seller whereby the protection buyer is compensated for the loss generated by a credit event (default of the reference entity, lack of payment of a coupon or other corporate events defined in the contract) in a reference credit instrument. In return, the protection buyer pays a premium, equal to an annual percentage *s* of the notional, to the protection seller. The premium *s*, quoted in basis points or percentage points of the notional, is called the CDS spread[2]. This spread is paid (semi)annually or quarterly in arrears until either maturity is reached or default occurs, at which point the protection seller pays the protection buyer the face value of the reference asset minus its post-default market value, through physical or cash settlement.

---

[1] The origin and initial purpose of CDS (invented by Blythe Masters, an executive at J.P. Morgan) was to exploit differences between the regulation of banks and insurance companies thus profiting from regulatory arbitrage, Kay (2016).

The economic costs and benefits of CDS contracts are the subject of intense debate. Extensive surveys are Augustin, Subrahmanyam, Tang, and Wang (2014) and Dias (2015). Anderson (2010) points out that although CDS may give social benefits in risk sharing and price discovery, these benefits may be undermined if the contract is prone to manipulation or if does not deal with counterparty risk. Heyde and Neyer (2010) posit that CDS trading increases the cyclicality of the banking sector in almost all economic scenarios because banks increase their investments in illiquid and risky assets and CDSs may create a channel of contagion.

However, the discussion in the literature about the role played by the CDS in the lending and securitization processes in the banking sector is scarce. Our main contribution in this paper is filling this gap. Specifically, we study the extent to which the mispricing in CDS contracts (and in securitized assets) affects the incentives of banks on securitization, borrowing and leverage and its implications for the cyclicality of the supply of bank credit and for the volatility of economic growth. We do so in a limits-to-arbitrage framework. The main message in this paper is that the mis valuation of credit-risk-sensitive financial assets has (mostly negative) effects in the real economy.

The key implications are as follows. Unlevered securitization together with mispricing of securitized assets boosts lending cyclicality, induces credit crunches and amplifies the cyclicality of the supply of bank credit, particularly when overpricing appears. If we include leverage, the volume of real activity and the profits of the banking sector increases, but banks sell securities when markets decline. This selling puts further pressure on falling prices.

---

[2] The 2009 CDS Big Bang (U.S.) and Small Bang (E.U.) protocols standardized premium payments. For instance, the fixed (yearly) premium payments for U.S. single-name CDS are either 50 or 100 basis points, and any difference relative to running par spread is settled through an upfront payment.

Regarding the economic impact of credit derivatives, trading in mispriced CDS reduces the amount of funding available to real-economy projects, in special those with high default risk. Trading in CDS decreases the liquidity of the securitized assets, and especially those based on investments with high credit risk. Banks consider buying naked CDS an attractive choice to lending to entrepreneurs in situations of financial stress when the cost of financing increases and counterparty risk is high. So, trading in CDS will decrease the amount of financing available to real-economy projects, so decreasing economic growth and employment.

Although we use a similar partial equilibrium model, our results are more general than the ones in Shleifer and Vishny (1997, 2010) because they do not consider securities with credit risk. We generalize their model by introducing credit-risk-sensitive assets and credit derivatives. In a paper related with ours, Heyde and Neyer (2010) show that CDSs reduce the stability of the banking sector both in recessions and booms because they incentive banks to increase their investment in risky credit portfolios and may create a possible channel of contagion. We extend their results by showing the interaction between leverage, securitization and, CDS and their impact on the credit supply cyclicality.

We organize the paper as follows. Section 2 reviews current literature. In Section 3 we introduce a limits-to-arbitrage partial equilibrium model. Section 4 summarizes the main results, discusses some empirical implications and suggests further research lines.

## 2. Literature Review

The possibility of CDS trading having effects at the aggregate level of the economy and the supply of bank credit has received attention in the literature. Although not many banks use CDS contracts (BIS, 2012), these banks are the bigger, the most systemic and the more interconnected. Thus, problems in the big banks could leak to other smaller banks and may affect the overall banking system. The economic costs of banking crisis can be sizeable (for instance, the cumulative output losses incurred during crisis periods are 15–20% of annual GDP, Hoggart, Reis, and Saporta, 2002) and so this issue is of paramount importance but has received little attention in the literature until recent times.

In a prescient paper, Allen and Carletti (2006) show credit risk transfer can lead to contagion and contribute to financial crises. Cont (2010) argues the impact of credit default swap markets on financial stability depends on clearing mechanisms and capital and liquidity requirements for large protection sellers. In particular, the culprits are not so much speculative or "naked" credit default swaps but inadequate risk management and supervision of protection sellers. When protection sellers are inadequately capitalized, CDS markets may act as channels for contagion and systemic risk. On the other hand, a CDS market where all major dealers take part in a central clearing facility with adequate reserves may contribute to mitigating systemic risk. Avellaneda and Cont (2010) suggest increased transparency in CDS markets benefits uninformed traders, while informed market participants (large dealers, market makers, and some large buy side firms) are likely to bear its costs.

There is controversy whether naked CDSs have played an important and direct role in destabilizing the financial system since 2007. Duffie (2010) argues naked trading on sovereign European CDS has not driven up Eurozone sovereign costs. He suggests that a regulation restricting speculative CDS trading[3] could reduce market liquidity, raise execution costs, lowering the quality of information provided by the CDS market and in the end increase sovereign borrowing costs. Heyde and Neyer (2010) posit CDS trading increases the cyclicality of the banking sector in almost all economic scenarios because banks increase their investments in illiquid and risky assets and CDSs may create a channel of contagion. The evidence in Das et al. (2014) suggests the advent of CDS was largely detrimental to firms because bond markets became less efficient and experienced no improvement in liquidity. Caglio et al. (2016) report that banks sell more credit protection than they buy for the firms in their loan and securities portfolios, so doubling the bet instead of hedging. This is the opposite of what we should expect if the reason to use CDS were to hedge credit risk.

---

[3] Such as the BaFin May 2010 ban in Germany on speculation against Eurozone sovereign bonds either through naked CDS or short bond positions

Mayordomo et al. (2014) discuss the inconsistency of CDS prices from different providers. The role of CDS in price discovery is analyzed in Forte and Peña (2009) and Mayordomo et al. (2011) analyze the impact of financial distress on the liquidity and the price discovery process.

Summing up, although some theoretical papers predict trading in CDS may increase financial instability, in the best of our knowledge, is scarce empirical evidence on the actual impact of CDSs on financial stability, economic growth, and systemic risk. Rodriguez-Moreno, et al. (2014) report , before the subprime crisis, holdings of credit derivatives by U.S. banks decreased systemic risk. However, after the crisis, these holdings increased banks' systemic risk. The impact of credit derivatives on systemic risk is significant but small. In fact, the proportion of non-performing loans to total loans and the leverage ratio have a much stronger impact on systemic risk than the level of holdings of credit derivatives. We should also stress that a key research question, the aggregate welfare effect of the CDS markets has received scant attention in the literature so far.

## 3. The Model

Given the scarce literature on the total impact of CDS on the cyclicality of bank credit, in this section, we present a limits-to-arbitrage model to analyze this problem. By the cyclicality of bank credit, we mean that there are economic reasons for cycles of credit expansion or contraction. Therefore, the cyclicality of bank credit exists when the funding of real-economy projects by banks concentrates in specific periods of time, but in other periods of time, there is almost no funding available for real-economy projects. Therefore, the cyclicality of bank credit is a crucial determinant of the volatility of real activity, output and employment.

We rely on three key assumptions. First, banks know the true default probability of a loan and this is private information that banks do not share with other agents. The justification for this assumption is the expertise banks should have in testing credit risk. Second, and given our first assumption, banks know whether market prices of securitized assets are above or below their fundamental values, and they use that information in their trading decisions. And third, the activity of the noise traders deviates market prices from fundamental values and banks, given our assumption of limits to arbitrage, are not able to put right at once and completely this market mispricing.

## 3.1. A Model of the cyclicality of bank credit: Loans, Securitization, Leverage and CDS

We work with a partial equilibrium model, with limits to arbitrage, based on the basic framework of Shleifer and Vishny (1997, 2010). However, we generalize the model by introducing credit risk and markets that allow credit risk transfer. Additionally, some suggestions in Gorton and Pennacchi (1995), Duffee and Zhou (2001), Heyde, and Neyer (2010) and Bolton and Oehmke (2011) are taken into account.

The model has three periods: 1, 2, and 3. The agents in the model are entrepreneurs and banks. Besides banks, other non-informed agents (noise traders) take part in the securitized loan market and in the credit risk transfer market. All agents are risk neutral and risk-free rates are zero. In this economy, there are various financial contracts: loans, securitized assets, short-term borrowing, and credit derivatives[4]. They can trade the loans in the market as securitized assets and we allow for temporary mispricing, caused by noisy traders, in the spirit of the limits-to-arbitrage literature. In our model, securitization is the sale of cash flow claims. The banks can leverage their investments by borrowing in the financial markets.

They can purchase the credit derivative instrument in the credit transfer market at a price $s$, pays off $w$ units (the loss given default of the loan) if the loan defaults at maturity and zero otherwise. We assume that the current practice of full collateralization of credit derivatives instruments[5] is the norm in the credit transfer market. We define the credit derivative instrument as a CDS because mimics the structure of existing CDS.

---

[4] We do not derive optimal financial contracts and simply assume that a reduced-form version of these contracts exists.
[5] See Augustin et al. (2014)

We study the consequences of the behavior of the different agents and the impact of the financial contracts for the cyclicality of lending decisions in the banking sector and the influence of this on the real economy.

### 3.1.1. Investment Projects

The entrepreneur presents for consideration investment projects to banks in periods 1 and 2 and all projects pay off in period 3. Each project costs €1 to undertake. There are two kinds of projects, good and bad. A project $Z$ starts at $t=1, 2$, and pays an amount $Z_i$ in $t=3$. This amount can be $Z_g \geq €1$ (if $Z$ is a good project) with probability $1 - \theta_Z$ or $Z_b = 0$ (if $Z$ is a bad project) with probability $\theta_Z$. The default probability of project $Z$ denoted as $\theta_Z$ and the project's payoffs are exogenously fixed and known by banks. The supply of projects costing €1 and yielding $Z_i$ is infinite, so their development only depends on the availability of funding.

The expected value of a project $Z$ at time $t=1, 2$ is then

$$E(Z) = (1-\theta_Z) Z_g + \theta_Z Z_b \quad (1)$$

We assume that entrepreneurs cannot get finance from the financial markets and therefore all projects are financed by banks[6]. When a bank finances a project (for instance the Z project) charges a fee of $f_Z$ to the entrepreneur at $t=1,2$ and receives from him a repayment $R_i$ $i=g,b$ at $t=3$. This repayment can be $R_g = €1$ or $R_b = 0$ in which case the entrepreneur defaults and the bank assumes the losses. The fee depends on the credit quality of the project:

$$f_K = g(\alpha, \theta_Z, Z_i) \quad (2)$$

with $0 \leq \alpha \leq 1$ depending on the competitiveness of the banking sector[7] and we require $\partial g / \partial \theta_Z > 0$, so the higher the project's default probability the higher the required proportional fee[8]. We assume that the entrepreneur pays the fee from his personal funds[9]. The simplest case is a situation in which the fee is proportional to the expected loss of the investment. Here $f_K = E*LGD*DP$ where $E$ is the exposure, $LGD$ is the loss given default and $DP$ is the default probability. In our case, this is equal to $f_Z = €1*1*\theta_Z$, that is, $f_Z = €\theta_Z$ [10].

---

[6] The major external source of finance for firms comes from banks and other financial intermediaries as suggested in Matthews and Thompson (2008). In our model entrepreneurs do not have previous reputation and the amount of capital they have plays no role in their financing.
[7] A perfectly competitive banking sector implies $\alpha = 0$. We also require $f_Z < 1$.

### 3.1.2. Banks

The representative bank starts the period $t=1$ with $E_0$ in equity. The government regulates the banking industry and sets compulsory capital requirements to safeguard financial stability. The bank faces a capital-asset ratio $e = E_t/N_t$ where $E_t$ is equity capital and $N_t$ is loaned to entrepreneurs. We define $N_t$ as the number of projects the bank finances at time $t =1,2$ and $E_t$ is the remaining bank's equity at the end of period $t =1,2,3$.

The bank can hold cash, invest in financial securities or lend money to entrepreneurs. We denote $C_t$ the amount of cash the bank holds at the end of period $t$ and we assume $C_2=C_3=0$ because there are no investment opportunities beyond time 2. To gain exposure to credit risk the bank can do four things. First, the bank may sell CDSs. Second, the bank may buy financial contracts based on securitized loans. Third, the bank may buy naked CDS. Finally, the bank may lend money to entrepreneurs for investment projects, in which case the bank collects the fee up front and receives the repayment $R_i$ at maturity $t = 3$. The bank can do three things with those loans. First, the bank can keep them on its books without hedging the credit risk. Second, the bank can securitize the loans and sell them in the market for securitized assets[11]. Finally, the bank can keep the loans on its books but hedging the credit risk by buying CDSs. Given that some loans are bad we assume that when the bank sells a loan in the market, the bank must keep a fraction $d$ of the loan on its books. We assume $d=h(\theta_Z)$ and $\partial h/\partial \theta_Z > 0$, so the higher the project's default probability the higher the fraction the bank keeps in its books. The fraction $d$ is named the "skin in the game" when the bank securitizes loans. If the bank finances $N$ projects, the bank must hold $dN$ of these securities on its balance sheet for at least one period after underwriting.[12]

---

[8] Assuming $\partial g/\partial Z_g > 0$ implies that the entrepreneur and the bank agree to share the expected surplus although this condition is not necessary for the main results in the paper.
[9] We assume that $f_Z$ is lower than €1 because otherwise the entrepreneur does not need banks to finance his project.
[10] Shleifer and Vishny (2010) assume that the entrepreneur and the bank split the surplus from the project, but in their model, there is no credit risk, in which case a linear parameterization of (2) is $f = \alpha(Z_g – 1)$.
[11] We do not consider packaging and tranching.

After securitizing a set of loans and keeping a fraction $d$ in its books, the bank sells the (1-$d$) remaining securities in the market. We denote by $P_{t,Z}$, $t=1,2,3$ the market price of the securitized assets based on project $Z$ at time $t$. Given that each securitized asset corresponds to a loan of €1 with a terminal value equal to $R_g$ or $R_b$ at $t=3$, its rational price must always be

$$P = P_{t,Z} = E(R_i) = (1-\theta_Z) \quad i=g,b \; ; \; t=1,2 \quad (3)$$

or in other words, the project's success probability. Market prices of securitized assets can deviate from (3) because of investor sentiment coming from a variety of sources unrelated to fundamental value. We talk about mispricing in more detail in section 3.1.3. below. We assume that banks understand the model, including that the fundamental value of the securitized asset is $P$. The bank has an incentive to securitize loans at $t = 1, 2$ only if $P_t > P$. If this is the case, the bank generate and sells loans at $t$, and collects profits, $B_t = (P_t - P)(1-d)$ per loan and distributes $(B_t+f)g$ in fixed proportions (e.g. 50/50) which are set exogenously as dividends and employee compensations (bonuses) where $g \leq 1$. Banking regulations can influence the part of retained profits (1- $g$) for instance setting countercyclical reserve provisions. Just as the bank sells assets when market prices are above the fundamental value, the bank may underpriced securities, that is when $P_t < P$. If this is the case, the bank buys loans at $t$, and collects profits $B_t = (P- P_t)$ per loan and distributes $(B_t)g$ as before. In both cases, banks' trading acts as a stabilizer of market prices but, given our assumption of limits to arbitrage, they are not able to put right the mispricing.

Recall that in this model, banks lend money to real-economy projects in periods 1 and 2. In the first period, banks lend a part $x$ of their funds to entrepreneurs and in the second

---

[12] This is the usual practice in the banking industry as documented in Gorton and Pennacchi (1995) who find that the most common arrangement is for the bank to retain a portion of the loan which is greater for riskier categories of loans

period banks lend a part $(1 – x)$. Full stability of bank credit is when $x = (1- x)$. Extreme cyclicality in bank credit is when $x = 0$ or $x = 1$.

Additionally, we address the question of how the bank finances its operations, besides its equity cushion. There is one source of financing in our model, short-term borrowing in the capital markets. The bank can borrow in financial markets using the securities it holds as a collateral $J_t$. We assume that the collateral can be sold at any time, so the bank's lenders see these loans as a safe investment. We denote by $L_t$ the stock of borrowing by the bank from the market at time $t=1, 2, 3$. Lenders to the bank want that the bank maintains a constant haircut $h$ in the form of securities on its debt and therefore $L_t = (1-h)J_t$.

In the following sections we first consider the case of $h=1$, so there is no short-term borrowing in the market and $w=0$ which implies that no credit risk protection is available. We compare baseline lending ($d=1$) with securitization ($d <1$). We ask when is profitable to the bank to finance everything at $t=1$ and save no cash for $t=2$, even when additional projects become available at $t=2$. Then we talk about the impact of retained profits and capital requirements, introduce leverage and we address the impact of introducing CDS contracts.

### 3.1.3. Mispricing

Market prices of securitized debt may deviate from the rational price (3) because of the investor's sentiment coming from a variety of sources (shifts in investor's psychology, regulatory rules, and market design) which are unrelated to fundamental payoffs[13]. For example, in the case of two equal securitized assets when one of them receives a rating from a Credit Rating Agency but the other does not. Given the current Basel regulatory framework which favors investment in securities with ratings, because unrated securities attract a high capital charge, most investors will prefer securities with a rating, increasing their prices in comparison with those of unrated securities. The deviation can be also in the opposite direction, for instance when unusual events cause investors to lose confidence in their valuation models and they react dumping securities (Caballero and Krishnamurthy, 2008). Market sentiment is especially prevalent in situations of Knightian uncertainty where information is too imprecise to be summarized adequately by probabilities. Informed agents (banks in our model) cannot put right market mispricing by a variety of reasons, such as noise trader risk or credit constraints. This gives rise to the phenomena of slow-moving capital to the investment opportunities, as emphasized by Duffie (2010).

---

[13] For a review on indicators of market sentiment see Baker and Wurgler (2007). The BW sentiment index is a contrarian predictor. In the case of loans and bonds they are probably like low sentiment beta stocks. In a period of high sentiment, they may be relatively underpriced and perform better than average going forward, and vice-versa.

Therefore, a key element in our model is the possibility that, given market dislocations, transactions costs, credit constraints, and other market imperfections, those frictions do not allow banks to cut out mispricing in the market and therefore market prices can deviate from fundamental values. For instance, with overpricing, potential buyers may not have enough financial muscle to buy the assets, there is synchronization risk (Abreu and Brunnermeier, 2002) and also the fact that potential buyers sometimes prefer to "ride the bubble" instead of correcting it (Brunnermeier and Nagel, 2004). This is the basic idea of the limits-to- arbitrage model (Shleifer and Vishny, 1997) which has received important empirical support, in the stock and bond market see Pontiff (1996), Baker and Savasoglu (2002), Lamont and Thaler (2003), Mitchell, Pulvino and Stafford (2002), Buraschi, Sener, and Menguturk (2011), and Mitchell and Pulvino (2012) among others[14]. Recently additional supportive evidence has been documented in the commodities market (Mou, 2011) and in the exchange rates market (Mancini and Ranaldo, 2011).

In our model mispricing appear because of the actions of some external (noise traders) investors whose demand affect security prices. Security prices depend on the assumption that arbitrage is limited and does not drive at once those prices to its fundamental value $P = E(R_i) = (1-\theta_Z)$ with securitized loans or $s = \theta_Z w$ with CDS. We assume that the banks understand that the fundamental values of securities are $P$ and $s$ at all times, but they do not share this knowledge with noise traders.

---

[14] For instance, Mitchell and Pulvino (2012) document many instances of persistent (more than three months) mispricing on the order of 10-20%. In some markets, they report far greater relative mispricing in the period 2007-2009. Buraschi et al (2011) report that in December 2008 Brazil's euro-denominated yield spread on 10-year Eurobonds was nearly 25% higher than the yield spread on the same maturity bond denominated in usd; this difference was only 4% in November 2005.

To clarify the mispricing process let's assume two agents: a bank and a noise trader and let's concentrate in period $t=1$. In period 1 the noise trader is pessimistic, optimistic or neutral about the value of security $i$, this sentiment is known by all at $t=1$ and it has size $\varphi_1 \geq 0$. When the noise trader is pessimistic[15], his period-1 demand schedule for security $i$ is given by:

$$QN_1 = (P - \varphi_1)/P_1 \quad (4)$$

Given that the bank has $E_1$ in equity and can also borrow in the capital markets the amount $L_1$ the demand schedule for the bank is:

$$QB_1 = (E_1+L_1)/P_1 \quad (5)$$

Since the total demand for the asset must equal the unit supply, the market price is given by:

$$P_1 = P - \varphi_1 + (E_1+L_1) \quad (6)$$

If bank's resources do not bring prices all the way to fundamentals, which implies $\varphi_1 > (E_1+L_1)$, then $P_1 < P$ or, in other words, the market price is below the fundamental price. Is easy to see that an overpricing appears when one optimistic shock cannot be compensated by short positions (because of lack of capital or lack of the suitable financial instrument) taken by banks in which case $P_1 > P$. The change in sentiment from period 1 to period 2 $\varphi_2 - \varphi_1$ is random and follows a given symmetric distribution $D$ with zero mean and constant volatility $\sigma$.

In summary, if banks (i) do not have access to enough debt capital, or (ii) cannot replace debt capital with new equity capital, or (iii) cannot take short positions, they cannot force the prices of assets to their fundamental values. Therefore, market frictions (e.g. credit constraints) limit arbitrage activities.

---

[15] We only consider the pessimistic sentiment; the optimistic sentiment is symmetric. The neutral sentiment is when $\varphi_1 = 0$

## 3.2. Lending versus Securitization

In this section, we deal in the first place with the case of traditional lending when there are no securitized assets, and market prices reflect fundamental values. In the second part of this section, besides traditional lending, banks can trade in securitized assets whose prices reflect fundamental values. In the third part, we deal with the case when market prices of securitized assets deviate from fundamental values.

### 3.2.1. Baseline Lending

We start with the baseline case in which the bank only engages in traditional lending activities and markets are informationally efficient. The projects available at $t=1,2$ are identical in the sense that the default probability $\theta_Z$ and the project's payoffs $Z_g$ and $Z_b$ are the same in every period[16]. Therefore the expected value of investing in $N$ projects is $E(Z_i)N$, $i=g,b$. If the bank uses all its balance sheet in $t=1$, lends out $E_0$ to finance $N=E_0$ projects and keeps all of them on its books until maturity. The bank collects $Nf$ as fees and distributes $(Nf)g$ among equity holders and employees. If the bank invests nothing in period 1 and invests all its balance sheet in period 2, collects the same fees which are distributed in the same way as before. Given that interest rates are zero the bank has no incentive to concentrate its investment in one given period. Thus, there is no reason for cycles of credit expansion or contraction. But, with identical projects, there is no reason justifying that the bank should smooth its financing over time. So, the situation may be of full stability (all bank financing evenly spread over periods), full instability (all bank financing concentrated in just one period) or anything between these extremes[17].

Financial regulations, however, can change that situation if different $g$ rates in different periods are required. For instance, if regulations say, at period $t=1$, that $g$ in period 1 will be lower than in period 2 (for instance because of a policy of dynamic provisioning), then the situation changes. On the one hand bank's equity holders are indifferent between receiving dividends in cash today or its equivalent value as capital gains at $t=3$. But a lower distribution ratio means that a higher amount of profits is keeping into the bank as reserves and will eventually increase capital gains[18]. Therefore, equity holders have an incentive to concentrate all the financing in period 1. The bank's managers and employees have a clear incentive to suggest shareholders skip the bank's investments altogether in period 1 and to invest in period 2 as much as possible. Whatever the way they resolve this agency conflict (favoring equity holders or managers) this situation would hint at some additional cyclicality in the framework of traditional lending.

---

[16] But in each period projects of different credit quality are available. For instance, a high quality project $X$ with $\theta_X = 0.01$ and a low quality project $Y$ with $\theta_Y = 0.90$.

[17] To circumvent this situation Shleifer and Vishny (2010) include some "special" high-payoffs projects available every period, but in limited supply, in order to incentive the bank to wait until period 2 and therefore smooth its lending between periods 1 and 2. If $N_H$ is the number of high-payoff projects available in each period, and if $E_0 \leq 2 N_H$, the bank finances $E_0/2$ projects each period and the situation is of complete stability.

### 3.2.2. Securitization

We turn to the case where the bank can securitize its loans. If the bank uses all its capital at $t=1$, the bank finances $N=E_0/d$ projects and keeps $dN$ securities in its books. Notice that given that $d<1$ then $N > E_0$ so the number of projects is higher than in the case with no securitization where $N=E_0$. The profits also increase with respect to the baseline case because the bank collects $Nf$ as fees and distributes $(Nf)g$ among equity holders and employees. If the bank uses all its capital in $t=2$, the result is the same. In fact, any combination of investments between periods 1 and 2 produces the same result. In summary, if market prices are consistent with fundamental values, we are in the same situation as in the baseline case, but with more investment projects, more economic activity, and higher bank profitability.

### 3.2.3. Mispricing

In this section, we talk about the impact of mispricing in securitized assets, caused by changes in the sentiment of noise traders, on the cyclicality of bank credit. Banks are the primary source of securitized assets because they are the originators of loans in our model. First, we analyze the implications on banks' behavior caused by market prices of securitized loans above their fundamental values (overpricing) and then we analyze the consequences of underpricing.

---

[18] For instance if distributed profits are allocated in 50/50 proportions to shareholders and employees and $g=0.2$ in $t=1$ and $g=0.4$ in $t=2$, equity holders will receive a total of $0.9(B+f)$ in dividends and capital gains if all investment is concentrated in $t=1$ but only $0.8(B+f)$ if all investment takes place in $t=2$

### 3.2.3.1. Overpricing

We first analyze the case when $P_1 > P$ and $P_2 = P$, so, in the first period, market prices of securitized loans are above their fundamental values, but this overvaluation disappears in the second period.

In this situation, the bank has a clear incentive to finance and securitize the maximum number of the investment projects the bank can afford at $t=1$ if the profits from selling the loans are superior to the cost of origination; that is when $P_1 - P > 1 - f$. If so, there are no incentives to wait for the next round of projects because banks do not have private information about future market price $P_2$. The reason is that this price is determined by a future change in the sentiment of the noise trader, which is, by assumption, unpredictable. This implies strong cyclicality in the lending process, in the sense that the bank invests all its capital in loans and later sells securitized assets if market prices overstate their true value, at $t=1$[19] and therefore, the bank collects fees and distributes profits among equity holders and employees at the end of period 1 and does nothing in period 2. This implies $x = 1$, a situation of full instability. Although selling activity by banks will depress prices at $t=1$, so mitigating the overpricing and helping to stabilize markets, recall that our assumption of limits to arbitrage prevents the full and immediate correction of mispricing.

In the second place, we analyze the case when $P_1 = P$ and $P_2 > P$, so, in the first period, market prices of securitized loans are at their fundamental values, but there is overvaluation in the second period. In this situation, the bank has no specific incentive for generating loans or distribute them at $t=1$, but, given that the bank does not know a future market opportunity will appear, the bank may invest all its capital at $t=1$, (i.e. $x = 1$) giving loans to entrepreneurs and keeping these loans in its books. If the bank does so, at $t=2$ the bank lends nothing to entrepreneurs and has a strong incentive to securitize all loans and sell them in the market, so decreasing (but not eliminating) the mispricing. If the bank does nothing at $t=1$ (i.e. $x = 0$) and waits until $t=2$, the bank has a clear incentive to finance and securitize all the investment projects at $t=2$, if the profits from selling the loans are superior to the cost of originating them; that is when $P_2 - P > 1 - f$. If so, there are no incentives to wait for the next round of projects because at $t=3$ there are no new projects. The bank will give loans and will sell securitized assets as long as market prices overstate their true value, at $t=2$. Although this selling activity by banks will depress prices at $t=2$, so mitigating the overpricing and helping to stabilize markets, recall that our assumption of limits to arbitrage prevents the correction of mispricing. Notice also that in both cases (either the bank lends everything in period 1 and sells securitized assets in period 2 or the bank does nothing at $t=1$ and waits until $t=2$ when the bank lends and securitizes) there is strong cyclicality in the lending process. A third situation is when the bank chooses $x = (1-x)$ as its investment strategy. In the first period the bank generates loans and may distribute them, but in the second period, the bank will sell all loans as securitized assets.

---

[19] Some evidence on the impact of securitization on housing bubbles can be found in Carbo-Valverde, Marques-Ibañez and Rodriguez-Fernandez (2012) who claim that mortgage-backed securitization together with housing prices were key factors in triggering the banking crisis in Spain.

In the third place, we analyze the case when $P_1 > P$ and $P_2 > P$. An additional point here is whether banks have perfect foresight and they know at $t=1$ that $P_2 > P$ (and as consequence, they know whether $P_1$ is higher or lower than $P_2$) as assumed in Shleifer and Visnhy (2010, page 309) or they do not have such abilities. If they have these abilities, then their optimal strategy is to concentrate all the originate-to-distribute activity in the most profitable point of time (either 1 or 2, i.e. $x = 1$, or $x = 0$) causing a situation of full instability. If they do not have such abilities, they will concentrate all the originate-to-distribute activity at time $t=1$, causing a situation of instability. Therefore, the optimal strategy of banks, despite their foresight abilities, is to concentrate all their lending in either period 1 or period 2 and securitize as much as they possibly can. In summary, in most cases, the overpricing in securitized assets creates strong incentives for lending cyclicality and as a result, there will be an increase in the volatility of output, and employment in the real economy. Deviations from fundamental values in (credit-risk-sensitive) securities markets impact the real economy through changes in the level of activity of the lending channel.

### 3.2.3.2. Underpricing

Next, we concentrate on the case in which $P_1 < P$ and $P_2 = P$, so, in the first period, securitized assets are undervalued. At $t=1$ the bank may generate loans, but the optimal strategy when $P-P_1 > 0$ is not to sell them. However, the bank does not know whether a future market mispricing may appear. Thus, the bank may invest all its capital at $t=1$ ($t=2$), giving loans to entrepreneurs and keeping these loans in its books. If the bank follows this strategy, at $t=2$ ($t=1$) the bank does nothing. Naturally, the bank can follow any intermediate strategy. But underpricing in $t=1$ discourages banks from entering the market for securitized assets, so they behave as if the market does not exist. Recall that banks are the only suppliers of securitized assets. So, they do not have incentives to trade and there are no counterforces to the market mispricing. In this situation, the strategy of shorting the undervalued asset hardly makes sense to banks, and noise traders do not generate loans or securitized assets. This situation is like the simple securitization situation, as discussed in section 3.2.2. In an additional scenario, at $t=1$ the banks lend a part of their capital to entrepreneurs, keep a part of their capital for investment in the next period, but do not sell securitized assets. In $t=2$, banks also finance investment projects, so there is little credit supply instability.

The results when $P_1 = P$ and $P_2 < P$, depend on the assumptions about the behavior of banks in period 1. Let's consider three possible scenarios. In the first scenario, at $t=1$ the banks lend all their capital to entrepreneurs. In $t=2$ banks do nothing because they have no funds to invest, so the underpricing does not affect their behavior, but entrepreneurs receive no financing during this period, so there is credit supply cyclicality. In the second scenario, at $t=1$ the banks lend a part of their capital to entrepreneurs, keep a part of their capital for investment in the next period and sell securitized assets to noise traders. In $t=2$, banks buy as much as they possibly can the underpriced securitized assets from noise traders and therefore, they are left with little money to finance investment projects, so there is also credit supply instability, although a version that is milder than in the first scenario. In the third scenario, banks do nothing in $t=1$, so there are no securitized assets available. In $t=2$ banks lend all their money to entrepreneurs but do not sell securitized assets. So, again, there is credit supply instability because they concentrate all lending in a single period.

When $P_1 < P$ and $P_2 < P$, let's consider three possible scenarios. In the first scenario, at $t=1$ the banks lend all their capital to entrepreneurs and do not securitize the loans. In $t=2$ banks do nothing because they have no funds to invest, so the underpricing does not affect their behavior, but entrepreneurs receive no financing during this period, so there is strong lending cyclicality. In the second scenario, at $t=1$ the banks lend a part of their capital to entrepreneurs, keep a part of their capital for investment in the next period, but do not sell securitized assets to noise traders, because there is underpricing. In $t=2$, banks cannot buy underpriced securitized assets from noise traders (they did not buy at $t=1$) and therefore they are left with money to finance investment projects, so there is little credit supply instability. In the third scenario, banks do nothing in $t=1$, so there are no securitized assets available. In $t=2$ banks lend all their money to entrepreneurs but do not sell securitized assets. So, again, strong credit supply cyclicality appears because they concentrate all lending in a single period.

In summary, although the situation is not as clear as was with overpricing, underpricing favors credit supply cyclicality in most cases.

Finally, we comment the case $P_1 > P$ and $P_2 < P$, overvaluation followed by undervaluation. Overvaluation incentives strong lending cyclicality. All investment and securitization concentrate in period 1 and therefore a credit crunch ensues in period 2[20].

Securitization raises the overall level of investment. But mispricing in the market for securitized assets also increases lending cyclicality. Besides, vagaries in the investor's sentiment (or changes in the regulatory framework) may influence the level of activity in the real economy through its relationship with the banking sector. This situation is more likely when securitized assets are overpriced.

**3.3. Retained Profits and Capital Requirements**

In this section, we talk about what happens when banking regulations impose $g < 1$ in $t=1$ and therefore cash reserves are available for investment in $t=2$. We assume that the bank granted loans and sold some of them to noise traders in $t=1$. We assume also $P_2 < P$. In this case, the income from financing new investment projects is $f$ for each project, while the profit from buying undervalued securities is $(P - P_2)$. The bank chooses the most profitable investment. Thus, the deeper the underpricing in $t=2$, the stronger the bank's incentive to buy distressed securities, at the expenses of direct lending to entrepreneurs. This situation favors lending cyclicality.

---
[20] The results in the case of $P_1 < P$ and $P_2 > P$ suggest an increase in banking instability.

The regulatory capital rules may generate strong incentives to securitize. For instance, if capital requirements on a class of loans are higher than other similar loans, banks will have an incentive to securitize that loan. With capital requirements with a capital-asset ratio $e = E_t/N_t$, an increase in $e$ increases the incentive to remove capital-intensive loans from the balance sheet which will lead to greater securitization. Similar to the case of reserve requirements, the regulations may force different $e$ ratios in different periods. For instance, if the regulations at the beginning of the period $t=1$ stablish that $e$ in period 1 will be lower than in period 2, banks have the incentive to originate and securitize in the latter period. This favors credit cyclicality. Summing up, securitization together with overpricing in the market of securitized assets encourages lending cyclicality and credit crunches. The provision of bank financing becomes more unstable. This situation may be further exacerbated by banking regulations such as dynamic provisioning.

### 3.4. Leverage

The bank can get additional funds borrowing short-term in the market and adjust to a haircut $h$. In this section, we concentrate on $P_1 = P$ and $P_2 < P$.

#### 3.4.1. Securitization with Short-term Borrowing

We assume that a bank which holds financial assets can borrow short-term in the market by using them as collateral. Lenders can always liquidate collateral if they want to be fully repaid and so, they want a zero interest rate. The mechanism to keep the borrowed money safe is the haircut $h$ that the bank must meet. When the bank uses all its resources in $t=1$ as collateral, then the haircut is set as

$$E_1/(E_1+L_1) = E_2/(E_2+L_2) = h \quad (7)$$

We first address the situation when at $t=1$ there is no market mispricing ($P_1 = P$) and the bank has equity $E_0$ and borrows $L_1=(1-h)J_1$. Without securitization, the bank uses the total amount $E_0+L_1$ to finance investment projects, so the total number of projects is $E_0+L_1=N$. With securitization, the bank can finance $N=(E_0+L_1)/d$ projects and keep $Nd$ securities on its books as skin in the game. The bank must maintain the haircut $h$ as required by its creditors and therefore $h= E_1/(E_1+L_1)$. Solving for the equilibrium number of projects we get that $N=E_1/dh$, and the collateral is $J_1= Nd$ which implies $E_0=E_1$ or in other words, the equity the banks have at the beginning and at the end of $t=1$ does not change because no change appears in the market values of the securitized debt. A bank with securitization but no leverage finances $N=E_0/d$ projects and with leverage finances $N=E_0/dh$ which is higher[21].

In the second period there is underpricing in the securitized assets market, i.e. $P_2 < P$. As shown in section 3.2.3.2., the optimal strategy when $P-P_2 > 0$ is always not to securitize. However, to maintain the haircut at the required levels the bank must sell securities. Consider the bank sells $S$ securities, so at the end of period 2, its collateral is $J_1 - S$ worth $(E_1/h-S)P_2$. The bank uses the proceeds from selling S securities to repay loans for an amount of $SP_2$, so the bank still owes $L_2 = L_1 - SP_2$. Is easy to see that the optimal amount to sell is $S=J_1Q$ where

$$Q=((1-P_2)/P_2)((1-h)/h) \qquad (8)$$

And where the impact on that part of the haircut on the market price is negative i.e. $dS/dh < 0$ $dS/dP_2 < 0$

The bank must liquidate a fraction $Q$ of its collateral. Notice that the two polar cases are $h=1$ (no liquidation) and $P_2 = 1-h$ (full liquidation). Remembering that $P=(1-\theta_Z)$ and that $P > P_2$ the implication is that $h > \theta_Z$ i.e. they set the haircut above the default probability. Abstracting from the extreme cases, falling prices and decreases in haircuts suggest faster portfolio liquidations.

---

[21] For instance, with $h=0.2$ and $d=0.2$ the bank finances 25 times its equity value when there is securitization and leverage but only 5 times its equity if only securitization is allowed.

Leverage changes the situation with respect to the simple securitization. First, the volume of real activity increases because the banks finance more projects. Second, banks sell securities when the prices of securitized assets decline. This selling puts further pressure on falling prices.

### 3.5. Credit Default Swaps

In this section, we talk about the consequences of introducing credit risk insurance (CDS) in the market for securitized loans and in the market for investment projects. The bank can buy or sell any amount of CDSs in the financial markets. The price of the CDS (spread) is *s*. The bank can buy or sell any amount of CDSs in the financial markets if the bank has enough collateral available. Given that the probability of default is $\theta_Z$, the zero-profit rule for all banks implies that the no-arbitrage CDS price is

$$s = \theta_Z * w \quad (9)$$

Whereas *w* is the loss given default, see Duffee and Zhou (2001).

We introduce here the concept of the CDS basis, we define as *s – f*. The CDS-bond basis for a maturity measures the difference between the credit default swap spread of a specific company and the credit spread paid on a bond of the same company, both instruments having the same maturity. In empirical studies, they usually define the basis as the difference between the price of the CDS and the price of the asset swap, which is a proxy for the conventional bond yield spread. Since the asset swap derives from a floating rate par bond, its price is more comparable to the CDS premium. The bond spread is the standard measure of corporate financing costs. In our model corporate financing costs are summarized by *f*. With no frictions in the corresponding markets, the CDS basis should be zero.

Technical and market factors (Choudry, 2006) favour a small positive basis in normal times. The factors favouring positive basis can be technical such as high-quality loans ($\theta_Z \approx 0$), non-standard credit events included in the CDS contract (CDS offers excess protection), cheapest to deliver option (CDS are more valuable), loans trading below par, funding costs below Libor and low counterparty risk. Additionally, there are market factors favouring positive basis: strong demand from protection buyers, liquidity premium (CDS is more liquid than a loan or a bond) and a shortage of cash assets among others. The empirical evidence on the size and sign of the basis[22] for corporate issuers suggests that, on average, in normal times is positive and small (about +5 bps for Investment Grade firms and +30 bps for High Yield firms). However, in crisis periods (e.g. 2007-2009), the basis becomes negative (about -250 bps for Investment Grade firms and -650 bps for High Yield firms). Reasons for a negative basis are (among others) strong increases in financing costs, reductions in liquidity and increases in the counterparty risk of the protection sellers.

### 3.5.1. Trading in CDSs versus Loans to Entrepreneurs

If the bank uses all its balance sheet in $t=1$ to lend to entrepreneurs, the bank uses all its equity $E_0$ to finance $N=E_0$ projects and keeps all of them on its books until maturity. For each loan, the bank's profits are $+f-1$ (loan origination) and $(1-\theta_Z)$ (expected payoff at maturity). Therefore, the total expected profit is $f - \theta_Z$.

If instead of giving loans to entrepreneurs, the bank gets exposure to credit risk by selling CDSs, the bank collects the CDS premium and sets aside €1 of its equity as collateral for each contract. The expected profit per contract is $s - \theta_Z$. The no arbitrage rule across markets implies that $s = f = \theta_Z$. Therefore, with fairly priced CDSs, the bank is indifferent between lending to entrepreneurs and selling CDSs, because in all cases the expected profit is zero.

Banks may decide between traditional lending and buying CDS when the basis is negative, but they prefer to sell CDS (with full collateralization) when the basis is positive. In the former case there is more investment and more economic growth if banks lend money to entrepreneurs, but no effect on the real economy if they buy CDS. If they sell CDS, there is less investment and less economic activity because the bank uses its funds as collateral to the CDS. In normal times the basis is positive and low for high quality creditors, and positive and moderate for less creditworthy reference

---
[22] Blanco, Brennan and Marsh (2005), De Wit (2006), Trapp (2009), Bai and Collin-Dufresne (2011)

names[23]. Therefore, in this normal situation, CDS affects lending to high-quality obligors but may restrict lending to less creditworthy projects[24]. If the basis becomes negative, no incentive appears to sell CDS. But is profitable to buy them because they are underpriced. This buying activity decreases the funds available for funding investment projects[25]. In summary, in most market situations, CDS reduces the amount of funding available to entrepreneurs and acts as a catalyst for a cut in lending, to high-risk borrowers.

### 3.5.2. Sales of CDSs versus Purchases of Securitized Assets

In what follows we analyze the case where $P_1 = P$ and $P_2 < P$. We assume that at $t=1$ the bank gives loans to entrepreneurs and sells securitized assets in the market. We also assume that the bank keeps the money to invest in period 2 (i.e. $0 < w < 1$). In t=2, the profits from buying underpriced securitized assets are $(P- P_2)$. The profit from selling CDSs is $s$. Comparing the profits from buying loans regarding the profits from selling CDSs it easy to see[26] that, for fairly priced CDSs, buying securitized assets or selling CDS are equally profitable with high-quality projects (default probability lower than 0.5). However, with riskier projects (default probability higher than 0.5), selling CDS is always preferable to buying securitized assets[27]. In these cases, we are in a situation where the loan market is decoupled from the CDS market, a situation that has been observed in times of severe financial distress, when default risk increases for most projects (Andriztky and Singh, 2007, Alexopoulou, Andersson, and Georgescu, 2009). In this scenario, selling CDS does not add liquidity to the securitized loan market, on the contrary, this selling activity decreases market liquidity. When the banks sell CDS protection, they go long in credit risk, equivalent to purchasing the loan. Does not make much economic sense buying loans, because that would double the position size.

---

[23] Trapp (2009) reports that for a sample of CDS, from June 1, 2001 to June 30, 2007, the average basis is 2.74 b.p. for investment grade bonds, but this average is 54.61 for non-investment grade bonds.
[24] This result is consistent with the empirical evidence in Ashcraft and Santos (2009). They find that, following the introduction of CDS in a sample of US firms, borrowing costs increase for low-credit-quality reference names, while these costs decreased for high-credit-quality reference names.
[25] The basis was significantly and persistently negative for most reference names during the GFC 2007-09, see Augustin et al. (2014)
[26] The profit from a loan purchase is $P-P_2$ and the profit from selling a CDS is $s$. Given the assumptions in the model, $P = (1 – \theta_Z)$ and fairly priced CDS implies s= $\theta_Z$. Equating both profit equations implies that the non-negative market price must be $P_2 = 1 - 2\theta_Z$ which implies $\theta_Z \leq 0.5$. If $\theta_Z > 0.5$, then $s > P-P_2$ because of the non-negativity constraint $P_2 \geq 0$.
[27] This is consistent with the evidence in Hirtle (2009) who suggests that the benefits of the growth of credit derivatives apply mainly to large and relatively safe firms.

Another situation when the bank has an incentive to sell CDSs instead of buying loans is when the capital rules set a high capital-asset ratio $e = E_t/N_t$, implying that buying loans requires increases in equity. Here, selling CDSs does not suffer from these problems.

In summary, mispricing in the securitized assets market implies that trading in CDS will decrease the liquidity of the securitized assets, and especially those based on investments exposed to high credit risk.

### 3.5.3. Trading in CDS versus Sales of Securitized Assets

The bank has an incentive to sell assets, i.e. to securitize loans at $t = 1, 2$ only if the difference between market price and fundamental value is profitable enough, i.e. when $P_t - P > 1 - f$. If this is the case, the bank sells loans and collects profits defined as $B_t = (P_t - P)(1-d)$ per loan. The bank can also consider keeping the loans in its balance sheet and buy credit insurance instead of securitizing. Notice that if the bank does so, the number of financed projects $N$ must be $E_0/(1+s)$ instead of $E_0$, so a lower number of projects get finance and there is an effect of reduced investment in the economy.

If the bank buys CDS, pays the spread $s$ and receives $R_i$ in period 3. For instance when $P_1 = P$ and $P_2 > P$ the profits from securitizing loans are $B_2 = (P_2 - P)(1-d)$ and the expected profit in $t=3$ is $E(B_3) = 0$. The profit from keeping the loan in the books and buying CDS is $B_2 = -s$ and the expected profit in $t=3$ is $E(B_3) = 1$ because either the loan does not default and the entrepreneur repays €1 or the loan defaults and then the protection seller[28] pays €1. Comparing the profits from securitizing loans regarding the profits from buying fairly priced CDS is easy to see[29] that, for high quality loans, which

---

[28] We assume no counterparty risk

[29] The profits from securitization are $(P_2 - P)(1-d)$ and the profits from buying CDS are $-s + 1$. Given the assumptions in the model, fairly priced CDS implies $s = \theta_Z$. Equating both profit equations gives the condition $(P_2 - P)(1-d) = 1 - \theta_Z$. For instance, in the case of a high quality loan where $d = 0.1$, and where

require small skin in the game and where the overpricing in the loan market is small, the bank prefers to keep the loans in the balance sheet and hedge their credit risk buying CDSs. Therefore, banks cut overall investment and economic activity slows down. large overpricing of bad quality loans tilts the balance towards securitization. The empirical implication of these results is that we should observe more demand for CDS protection in the case where the reference entity is of high quality and trades in a liquid market with a small basis. We should observe more securitization activities with low-quality loans traded in illiquid or informationally inefficient markets.

### 3.5.4. Purchases of Naked CDS

This is a speculative strategy which implies using the entire bank's equity as payments for buying credit protection and they finance no investment projects, which implies an immediate negative effect on the volume of investment in the economy. The number of CDS contracts bought is equal to $E_0/s$, and given that the spread $s$ is lower than one, the notional amount of protection bought is much larger than the number of projects that the bank would finance if the bank lends to entrepreneurs[30].

When the bank buys naked CDS, pays the spread $s$ and receives $R_i$ in period 3. Suppose we are in $t=2$. The cost of buying one CDS is $s$ and the expected profit is the payoff €1 times the default probability $\theta_Z$, that is $\theta_Z$. This situation of a negative basis may appear when the funding costs explode over Libor or if the counterparty risk in the CDS market increases, among other markets factors (e.g. liquidity crunches). This has been the case in the crisis period 2007-2009.

These situations (steeper financing costs, high counterparty risk, and low liquidity) are usually associated with episodes of market stress. Is in those periods when buying naked CDSs become more attractive to banks. The economic consequences are a cut in investment in the real economy. Therefore, in situations of financial distress, trading in CDS increases credit restrictions, credit supply instability, and underinvestment.

---

$(P_2 - P) = 0.1$ and $\theta_Z = 0.2$ it is clearly preferable to keep the loan and hedge its credit risk with a fairly priced CDS.

[30] Norden and Radoeva (2013) study the CDS volume on a firm that exceeds its outstanding debt. This naked CDS volume indicates speculation since hedging can be ruled out. They provide evidence of substantial speculation in the CDS market. The mean ratio of the firm-specific CDS volume to total debt is 3.6 and the maximum is 65.

## 4. Summary and Conclusions

The conclusions from this paper give a nuanced answer to whether CDSs are helpful or harmful in aggregate economic terms. If the markets for securitized assets and credit risk protection are informatively efficient, the CDS are redundant securities with very small impact on the cyclicality of the credit supply and on the volatility of economic growth.

The mispricing of assets traded in these markets creates incentives for banks to engage in trading. This trading has effects on the real economy through its impact on the lending channel because banks engage in trading using funds which, in normal circumstances, should be invested in projects of the real economy. Therefore, mispricing of (credit-risk-sensitive) financial assets have effects in the real economy. Our main conclusions are

1) Unlevered securitization together with mispricing of securitized assets boosts lending cyclicality, particularly when overpricing appears.
2) Levered securitization increases the volume of real activity and the profits of banks, but banks sell securities when markets decline. This selling exacerbates downward movements in the securitized assets market.
3) With mispriced CDS or securitized assets, trading in CDS reduces the amount of funding available to real-economy projects, in special those with high default risk.
4) Trading in CDS decreases the liquidity of the securitized assets, and especially the liquidity of assets based on investments with high credit risk.
5) Trading in naked CDS decreases the amount of financing available to real-economy projects, especially in situations of economy-wide financial stress.

Some empirical implications of the above results are worth noting. First, when the CDS basis is close to zero the impact of CDS on lending cyclicality, securitization, and economic activity is unlikely to be very relevant. However, the higher (in absolute value) the basis, the stronger the CDS impact is likely to be on lending cyclicality, securitization, and economic activity. Second, the impact of CDS should be stronger if the credit quality of the reference name is relatively low. Third, we should see an increase in the relative volume in the CDS market, in comparison with the lending and securitization markets, when financial stress increases across the economy.

Looking forward, introducing heterogeneous agents in the model may yield more insights. Allowing for uncertainty in banks' information set may give more general results. For instance, uncertainty of the banks' knowledge of whether an asset is mispriced. Asset prices may also be endogenous and dependent on the economic cycle. A generalization of the model allowing for a welfare analysis including externalities and the specific social benefits and costs of CDS trading is an interesting research area. An important aspect, which our model does not consider, is that banks can use CDS for other purposes (e.g. regulatory capital relief) that can incentivize its use, see Yorulmazer (2013) and the empirical evidence on European banks in Thornton and di Tommaso (2018) suggesting that CDS are used for regulatory arbitrage to lower capital rules and boost risk taking. We leave all these topics for future research.